\documentstyle[aps,12pt,manuscript]{revtex}
\begin{document}
\preprint{INJE-TP-99-3}

\title{Gauge bosons and the AdS$_3$/LCFT$_2$ correspondence}

\author{ Y. S. Myung and H. W. Lee }
\address{Department of Physics, Inje University, Kimhae 621-749, Korea} 

\maketitle

\begin{abstract}
We study the relationship between the gauge boson coupled to 
spin 2 operator and the singleton 
in three-dimensional anti-de Sitter space(AdS$_3$).
The singleton can be expressed in terms of a pair of dipole ghost fields 
$A$ and $B$ which couple to $D$ and $C$ operators 
on the boundary of AdS$_3$.
These operators form the logarithmic conformal field theory(LCFT$_2$). 
Using the correlation function for logarithmic pair, 
we calculate the greybody factor for the singleton.
In the low temperature limit of $\omega \gg T_{\pm}$, this 
is compared with the result of the bulk AdS$_3$ calculation 
of the gauge boson.
We find that the gauge boson cannot be realized as a model of 
the AdS$_3$/LCFT$_2$ correspondence.
\end{abstract}
%\vfill
%Compiled at \today : \number \time. 

\newpage
Recently the AdS/CFT correspondence has attracted much 
interest\cite{Mal98ATMP231,Gub98PLB105,Wit98ATMP253}.
This means that the string/M theory on AdS$_{d+1}$ is dual to 
the gauge theory of the CFT$_d$ on its boundary.
This was used to resolve many problems in black hole 
physics\cite{Gub98NPB393,Tay9806132,Teo98PLB269,Mul9809193}.
For the test of the AdS$_3$/CFT$_2$ correspondence one introduces a set of 
test fields \{$\Phi_i$\} on AdS$_3$ and their corresponding 
operators \{${\cal O}^i$\} on the boundary.
For example, these are in the D1-D5 brane system : 
a free scalar($\phi$) which couples to (1,1) operator; 
two fixed scalars($\nu, \lambda$) to (2,2), (3,1), and (1,3) operators;
two intermediate scalars($\eta, \xi$) to (1,2) and (2,1) 
operators\cite{Cal97NPB65}.
The relevant relation between these is given by\cite{Wit98ATMP253}
\begin{equation}
e^{-S_{\rm eff}(\{\Phi_i\})} =
\langle e^{\int_{\rm B} \Phi_{b,i} {\cal O}^i } \rangle.
\label{bulk-boundary}
\end{equation}
The expectation value $\langle \cdots \rangle$ is taken in the CFT 
with the boundary test field $\Phi_{b,i}$ as a source.
Eq.(\ref{bulk-boundary}) was widely used for calculations of the entropy, 
greybody factor(2-point function), 3- and 4-point functions.
It was shown by studying exchange diagram with scalar and gauge fields 
in N=4 SUSY Yang-Mills from AdS$_5$ that the 4-point function 
has logarithmic singularities\cite{Fre98strings}. 
More recently Kogan proposed that a dipole ghost pair($A,B$) 
can represent a singleton, which  
induces the 2-point correlation function for a logarithmic 
pair(${\cal O}, {\cal O}'$)\cite{Kog9903162}.
He argued that this is the origin of logarithmic singularities 
in the 4-point functions.
This logarithmic pair with the normalization factor 
$c=2 (\Delta-1)^2/\pi$\cite{Teo98PLB269} has the 2-point 
correlation fucntions\cite{Ghe9807034,Gur93NPB535}
\begin{eqnarray}
\langle {\cal O}({\bf x}) {\cal O}'({\bf y}) \rangle &=& 
      { c \over {|{\bf x}-{\bf y}|^{2 \Delta}}},
\nonumber \\
\langle {\cal O}'({\bf x}) {\cal O}'({\bf y}) \rangle &=& 
      { c \over {|{\bf x}-{\bf y}|^{2 \Delta}}} \left [ 
              -2 \ln|{\bf x}-{\bf y}| +  { 1 \over c} 
              {\partial c \over \partial \Delta} \right ],
\label{2-point} \\
\langle {\cal O}({\bf x}) {\cal O}({\bf y}) \rangle &=& 0.
\nonumber 
\end{eqnarray}
Here $\Delta$ is a degenerate dimension $\Delta$ of 
${\cal O}$ and ${\cal O}'$ and we note a crucial relation 
$\langle {\cal O}'({\bf x}) {\cal O}'({\bf y}) \rangle = 
{\partial \langle {\cal O}({\bf x}) {\cal O}'({\bf y}) \rangle
\over \partial \Delta}$.
We will use this relation to calculate the greybody factor.

In the AdS$_3$/CFT$_2$ correspondence, there exists a puzzle of the 
missing states between CFT$_2$ and supergravity\cite{Vaf9804172}. 
The gauge bosons appear in the resolution of this puzzle.
These are chiral primaries which correspond to the descendent of the 
identity operator in the CFT$_2$\cite{Mal9804085}.
But on the supergravity side these are absent and thus 
may be considered as unphysical singletons on 
AdS$_3$\cite{Boe9806104}.
In this sense, it is important to test the 
relationship between gauge boson and singleton.
The authors in \cite{Kim9812016} found that these gauge bosons coupled 
to (2,0) and (0,2) 
operators on the boundary receive logarithmic corrections from 
a bulk AdS$_3$ scattering calculation.
Of course this was performed in the low temperature limit of
$\omega \gg T_{\pm}$.

In this paper, we derive the greybody factor for the singleton. 
According to Ref. \cite{Kog9903162}, 
we wish to represent this with a pair of dipole ghost fields($A,B$) 
coupled to (1,1) operators($D,C$).
This means that a gauge boson with spin 2 
assumed to be expressed in terms of 
a pair of dipole ghost fields with spin 0.
And then we calculate the two-point function of their 
operators in terms of the BTZ coordinates.
Using this, we obtain the greybody factor.
Finally we will compare this with 
the result of a bulk AdS$_3$ scattering in the low-temperature limit.

We start with the bulk AdS$_3$ action for a dipole 
pair($A,B$)\cite{Ghe9807034,Kog9903162}
\begin{equation}
S_{\rm eff} = \int d^3 x \sqrt{g} \left [ \partial A \cdot \partial B 
   - m^2 A B - {1 \over 2} B^2 \right ].
\label{dipole-action}
\end{equation}
At this stage it is not clear if this action comes from 
supergravity(string) theories.
Rather, (\ref{dipole-action}) takes a similar form of the 
Nakanishi-Lautrup formalism in the gauge theory\cite{Kug79PTPS1}.
In detail, (\ref{dipole-action}) with $m^2 = 0$ and $A_\mu = \partial_\mu A$ 
leads to the gauge-fixing term as 
$S_{\rm GF} = - \int d^3 x \sqrt{g} \left [ B \partial_\mu A^\mu 
+ { \alpha \over 2} B^2 \right ] $ with $\alpha=1$. 
Here $B$ is the Nakanishi-Lautrup field.
$A$ corresponds to $\sigma$ in \cite{Kug79PTPS1} and 
leads to the negative-norm state.
Its equations of motion are given by
\begin{equation}
\left ( \nabla^2 + m^2 \right ) A + B =0, ~~
\left ( \nabla^2 + m^2 \right ) B =0 .
\label{dipole-equation}
\end{equation}
A solution to these can be found from a 
boundary-bulk Green function for $\langle AB \rangle$ with mass 
$m$\cite{Kes9808037,Mul9809193}
\begin{eqnarray}
&&K_{AB}(r, u_+, u_-; u_+', u_-') 
=K_{BA}(r, u_+, u_-; u_+', u_-') 
\nonumber \\
&&~~~~~~~~~~~~~~= N \left \{
{\pi^2 T_+ T_- \over 
{ {{r_+^2 - r_-^2} \over 4r } \exp(\pi [ T_+ \delta u_+ + T_-\delta u_- ])
+ r \sinh(\pi T_+\delta u_+)\sinh(\pi T_-\delta u_-) } }
\right \}^\Delta,
\label{green-function}
\end{eqnarray}
where we used the coordinates ($r,t,\phi$) in the BTZ black hole with
$R=1$ : mass $M=(r_+^2 - r_-^2)$, angular momentum $J=2 r_+ r_-$,
and left/right temperatures 
$T_\pm = (r_+ \pm r_-)/2 \pi $\cite{Ban93PRD1506}.
The Hawking temperature $T_H$ is defined by 
$2/T_H = (1/T_+ + 1 /T_-)$.
Here $u_\pm= \phi \pm t, \delta u_\pm = u_\pm - u_\pm'$, 
$\Delta(\Delta-2)=m^2$.
The normalization constant $N$ is introduced for convenience.
To find $K_{AA}$ we have to solve the equation
\begin{equation}
\left ( \nabla^2 + m^2 \right ) K_{AA} = - K_{AB}.
\label{KAA-equation}
\end{equation}
This is found by a trick as 
\begin{equation}
K_{AA} = { \partial K_{AB} \over \partial m^2 } =
  {1 \over 2 (\Delta -1)} { \partial K \over \partial \Delta }.
\label{KAA-solution}
\end{equation}
Then the solution is given by 
\begin{eqnarray}
A(r, u_+, u_-) =&& \int d u_+' du_-' 
\left [ K_{AB}(r, u_+, u_-; u_+', u_-') B_b(u_+', u_-') + \right .
\nonumber \\
&&~~~~~~~~~~~~~~~~~~~\left . 
K_{AA}(r, u_+, u_-; u_+', u_-') A_b(u_+', u_-')\right ],
\label{AB-solution} \\
B(r, u_+, u_-) =&& \int d u_+' du_-' 
 K_{BA}(r, u_+, u_-; u_+', u_-') A_b(u_+', u_-') ,
\nonumber 
\end{eqnarray}
where $A_b$ and $B_b$ are the boundary values of $A$ and $B$ respectively.
Using (\ref{dipole-action}) and (\ref{dipole-equation}), the 
effective action takes only the boundary form
\begin{equation}
S_{\rm eff}[A_b, B_b] = \lim_{r_s \to \infty} {1 \over 2} \int_{S} du_+ du_-
  \sqrt{-h} \left \{ A ( \hat n \cdot \nabla ) B +
         B ( \hat n \cdot \nabla ) A \right \},
\label{effective-action}
\end{equation}
where $S$ is a regularized surface at $r=r_s$ and 
$\hat n \cdot \nabla = r (\partial / \partial r)$.
$h_{\mu \nu}$ is an induced boundary metric with 
diag($-r^2, r^2$) and thus $\sqrt{-h} = r^2$.
Considering the boundary behavior of $A$ and $B$ as 
\begin{equation}
A(B)|_{r \to \infty} \sim r^{-2 + \Delta} A_b(B_b),
\label{A-boundary}
\end{equation}
one finds 
\begin{eqnarray}
S_{\rm eff}[A_b, B_b] = && - { \Delta N \over 2 } 
\int du_+ du_- du_+' du_-'
\left [ { \pi T_+ \over \sinh(\pi T_+\delta u_+) } \right ]^\Delta
\left [ { \pi T_- \over \sinh(\pi T_-\delta u_-) } \right ]^\Delta
\nonumber \\
&& \times
\Bigg [ 2 A_b(u_+, u_-) B_b(u_+', u_-') 
\label{action-boundary} \\
&& ~~ \left .
+ {A_b(u_+, u_-) A_b(u_+', u_-') \over 2 (\Delta -1) }
\left \{ {1 \over N} {\partial N \over \partial \Delta} - \ln \left (
{ \sinh(\pi T_+\delta u_+) \over \pi T_+} \cdot
{ \sinh(\pi T_-\delta u_-) \over \pi T_-} \right ) 
\right \} \right ].
\nonumber
\end{eqnarray}
With (\ref{bulk-boundary}), one can derive the two-point functions 
for conformal operators $C$ and $D$ as
\begin{eqnarray}
\langle C(u_+, u_-) C(0) \rangle &=&
-{\delta^2 S[A_b, B_b] \over \delta B_b(u_+, u_-) \delta B_b(0) } =0 ,
\label{conformal-CC}\\
\alpha \langle C(u_+, u_-) D(0) \rangle &=&
-{\delta^2 S[A_b, B_b] \over \delta B_b(u_+, u_-) \delta A_b(0) } 
\nonumber \\
&=&
{\Delta N } 
\left [ {\pi T_+ \over \sinh(\pi T_+ u_+) } \right ]^\Delta
\left [ {\pi T_- \over \sinh(\pi T_- u_-) } \right ]^\Delta,
\label{conformal-CD}\\
\beta \langle D(u_+, u_-) D(0) \rangle &=&
-{\delta^2 S[A_b, B_b] \over \delta A_b(u_+, u_-) \delta A_b(0) } 
\nonumber \\
&=&
{N \Delta \over 4(\Delta-1)} 
\left [ {\pi T_+ \over \sinh(\pi T_+ u_+) } \right ]^\Delta
\left [ {\pi T_- \over \sinh(\pi T_- u_-) } \right ]^\Delta
\nonumber \\
&&~~~~~~ \times
\left [ {1 \over N} {\partial N \over \partial \Delta } - \ln \left \{
{\sinh(\pi T_+ u_+) \over \pi T_+ } \cdot
{\sinh(\pi T_- u_-) \over \pi T_- } \right \}
 \right ],
\label{conformal-DD}
\end{eqnarray}
where $\alpha, N $, and $\beta$ are chosen to recover (\ref{2-point})
in the low temperature limit of $T_\pm \to 0$.
Then one finds $\alpha =\Delta$, $N=c, \beta=\Delta/4 (\Delta -1)$.
We are now in a position to calculate the greybody factor using 
the above correlation functions\cite{Mal97PRD4975}.
The greybody factor for $\langle AB \rangle$ is calculated by the boundary CFT
as\cite{Gub97PRD7854,Teo98PLB269}
\begin{eqnarray}
\sigma_{\rm abs}^{AB} &=& { \pi \over \omega} \int dt 
\int_{-\infty}^\infty d \phi e^{-i \omega t + i p \phi} 
\left [ \langle C(t - i\epsilon, \phi) D(0) \rangle -
        \langle C(t + i\epsilon, \phi) D(0) \rangle \right ]
\nonumber \\
&=&
{2(\Delta-1)^2 (2 \pi T_+R)^{\Delta -1} (2 \pi T_- R)^{\Delta -1}
    \sinh({\omega \over 2 T_H}) \over 
\omega \Gamma^2(\Delta) }
\nonumber \\
&&~~~~~~~~~~~~~~~~~~~~~~~~~~\times
\left \vert \Gamma \left ({\Delta \over 2} + i {\omega \over 4 \pi T_+} \right )
          \Gamma\left ( {\Delta \over 2} + i {\omega \over 4 \pi T_-} \right )
\right \vert^2,
\label{abs-CD}
\end{eqnarray}
where $R^{2(\Delta -1)}$ has been switched on $\langle CD \rangle$, 
to recover a complete form of the greybody factor.
Here the original integral region of $0 \le \phi \le 2 \pi$ is changed 
into $-\infty \le \phi \le \infty$ to accommodate the periodic nature of
$u_\pm \sim u_\pm + 2 \pi n$ for the BTZ black hole\cite{Mul9809193}.
For $\Delta=2(m^2=0)$, (\ref{abs-CD}) takes exactly the same form of the greybody 
factor for a massless minimally coupled 
scalar($\nabla^2 \Phi = 0$)\cite{Bir97PLB281}
\begin{equation}
\sigma_{\rm abs}^{AB} = \pi^2 \omega R^2 
{e^{\omega/T_H} -1 \over \left ( e^{\omega/2 T_+} - 1 \right ) 
\left ( e^{\omega/2 T_-} - 1 \right )}.
\label{abs-minimal}
\end{equation}
In the low energy limit of $\omega \ll T_\pm$ one finds 
$\sigma_{\rm abs}^{AB} \vert_{\omega < T_\pm} =$ $ 2 \pi r_+ = $
${\cal A}_H$, while in the low temperature limit of $\omega \gg T_\pm$ 
it takes $\sigma_{\rm abs}^{AB} \vert_{\omega > T_\pm} =$ $\pi^2 \omega R^2$.

It seems to be difficult to calculate the greybody 
factor $\sigma_{\rm abs}^{AA}$ directly, because of the 
logarithmic singularities in (\ref{conformal-DD}).
Instead, we use the relation of 
$\langle D(u_+,u_-) D(0) \rangle$ $= {\partial \over \partial \Delta}
\langle D(u_+,u_-) C(0) \rangle$ and thus expect to find 
$\sigma_{\rm abs}^{AA} \simeq \partial \sigma_{\rm abs}^{AB}/ \partial \Delta$.
In this calculation we have to use the relation for the gamma function 
as
\begin{equation}
{\partial \Gamma(z) \over \partial \Delta} =
{\partial z \over \partial \Delta} {\partial \Gamma(z) \over \partial z}
= {\partial z \over \partial \Delta} \Gamma(z) \psi(z),
\label{gamma-relation}
\end{equation}
where $\psi(z) = \partial \ln \Gamma(z) / \partial z$ is a digamma function.
Finally we obtain the greybody factor for 
$\langle AA \rangle$ by using the boundary LCFT$_2$ 
correlator $\langle DD \rangle$ as
\begin{eqnarray}
\sigma_{\rm abs}^{AA} &=& { \pi \over \omega} \int dt 
\int_{-\infty}^\infty d \phi e^{-i \omega t + i p \phi} 
\left [ \langle D(t - i\epsilon, \phi) D(0) \rangle -
        \langle D(t + i\epsilon, \phi) D(0) \rangle \right ]
\nonumber \\
&\simeq&
{2 (\Delta-1)^2 (2 \pi T_+R)^{\Delta -1} (2 \pi T_- R)^{\Delta -1}
    \sinh({\omega \over 2 T_H}) \over 
\omega \Gamma^2(\Delta) }
\label{abs-DD} \\
&&\times
\left \vert \Gamma\left ( {\Delta \over 2} + i {\omega \over 4 \pi T_+} \right )
          \Gamma\left ( {\Delta \over 2} + i {\omega \over 4 \pi T_-} \right )
\right \vert^2
%\nonumber \\
%&&\times
\Bigg [
{2 \over \Delta-1 } + \ln(2 \pi T_+R) + \ln(2 \pi T_-R) - 2 \psi(\Delta)
\nonumber \\
&&~~~~+{1 \over 2} \left \{
\psi({\Delta \over 2}+i{\omega \over 4 \pi T_+}) +
\psi({\Delta \over 2}-i{\omega \over 4 \pi T_+}) +
\psi({\Delta \over 2}+i{\omega \over 4 \pi T_-}) +
\psi({\Delta \over 2}-i{\omega \over 4 \pi T_-}) 
\right \}
\Bigg ].
\nonumber 
\end{eqnarray}
As far as we know, this is the first result for a dipole ghost pair(singleton).
Now let us compare this with the result of a gauge boson 
with $s=2$ from a bulk AdS$_3$ scattering\cite{Kim9812016} 
\begin{equation}
\sigma_{\rm abs}^{\rm gb} = \pi^2 \omega R^2 s^2 [ 1 + \omega R s 
\ln(2 \omega R s) ].
\label{abs-AdS}
\end{equation}
For this purpose, we take the low temperature limit of
$\omega \gg T_\pm$ and $\Delta=2$ on $\sigma_{\rm abs}^{AA}$.
In this case the digamma function $\psi$ can take an asymptotic 
form as\cite{Abr66}
\begin{equation}
\Re \psi(1+iy) = \Re \psi(1-iy) \simeq 
\ln y + {1 \over 12 y^2} + {1 \over 120 y^4} + \cdots.
\label{digamma-series}
\end{equation}
In the low temperature limit, $\sigma_{\rm abs}^{AA}$ takes the form
\begin{equation}
\sigma_{\rm abs}^{AA} = { \pi^2 \omega R^2 } \left [
           1 + 2 \ln(\omega R) + c' \right ],
\label{abs-final}
\end{equation}
where $c' = 2 \gamma -1 -2 \ln2$ with the Euler's constant $\gamma=0.5772$.
At the first sight, it seems that 
the (\ref{abs-final}) 
takes a similar form as in (\ref{abs-AdS}).
In (\ref{abs-AdS}) the logarithmic term is multiplied by $\omega R$ and 
thus it is a subleading-order.
However, in (\ref{abs-final}) one cannot find such a prefactor and 
$2 \ln(\omega R)$ is regarded as the leading low energy behavior.
In this sense, we argue that there is no agreement between 
the AdS$_3$ calculation 
of gauge field and the LCFT result of a singleton.
Further, the non-logarithmic terms in (\ref{abs-final}) do not agree 
with that of (\ref{abs-AdS}).
If the spin 2 gauge boson is truely represented by a pair of dipole 
ghost fields on AdS$_3$, from the AdS/LCFT correspondence 
(\ref{abs-AdS}) should agree with (\ref{abs-final}) 
even in the low-temperatue limit.
Hence we conclude that the gauge boson with spin 2 has nothing to do with 
the AdS/LCFT correspondence.

On the other hand one finds the logarithmic operators in (\ref{conformal-DD}), 
which may induce the unitarity problem.
Here we may resolve this problem.
It is noted that these logarithmic terms originate from the 
unphysical dipole ghost fields ($A,B$).
As was shown in \cite{Kug79PTPS1}, this pair 
($A,B$) is turned into the zero-norm state by the Goldstone dipole 
mechanism in Minkowski spacetime.
We suggest that the boundary logarithmic terms come from 
the negative-norm state of $A$.
In order to remove the negative-norm state, 
we impose the subsidiary condition as $B^+(x) \vert 0 \rangle_{\rm phys}=0$.
Then the physical space($\vert 0 \rangle_{\rm phys}$) 
will not include any $A$-particle state.
This corresponds to the dipole mechanism to cancel the negative-norm state.
Similarily, we expect that in the boundary CFT$_2$ of AdS$_3$, 
the theory can be managed to be unitary 
by choosing an appropriate subsidiary condition.

In conclusion, we derive the new greybody factor for a singleton 
from the LCFT$_2$ correlator $\langle DD \rangle$
which corresponds to the derivative of the CFT$_2$ 
correlator $\langle DC \rangle$ with respect to the weight $\Delta$.
In the low temperature limit the bulk AdS$_3$ result 
$\sigma_{\rm abs}^{\rm gb}$ does not lead to $\sigma_{\rm abs}^{AA}$ of 
the LCFT$_2$ correctly.
This means that the spin 2 gauge boson 
cannot be expressed in terms of a pair of dipole ghost fields ($A,B$). 

\acknowledgments
This work was supported in part by the Basic Science Research Institute 
Program, Ministry of Education, Project NOs. BSRI-98-2413 and BSRI-98-2441.


\begin{references}
\bibitem{Mal98ATMP231}
  J. Maldacena, Adv. Theor. Math. Phys. {\bf 2}, 231(1998), hep-th/9711200.
\bibitem{Gub98PLB105}
  S.S. Gubser, I.R. Klebanov and A.M. Polyakov,
        Phys. Lett. {\bf B428}, 105(1998), hep-th/9802109.
\bibitem{Wit98ATMP253}
  E. Witten, Adv. Theor. Math. Phys. {\bf 2}, 253(1998), hep-th/9802150.
\bibitem{Gub98NPB393}
  S. Gubser, A. Hashimoto, I. Klebanov and M Krasnitz,
          Nucl. Phys. {\bf B526}, 393(1998), hep-th/9803023.
\bibitem{Tay9806132}
  M. Taylor-Robinson, hep-th/9806132.
\bibitem{Mal9804085}
  J. Maldacena and A. Strominger, hep-th/9804085.
\bibitem{Teo98PLB269}
  E. Teo, Phys. Lett. {\bf B436}, 269(1998), hep-th/9805014.
\bibitem{Mul9809193}
  H. M\"uller-Kirsten, N. Ohta, and J. Zhou, hep-th/9809193.
\bibitem{Cal97NPB65}
  C. Callan, S. Gubser, I. Klebanov, and A. Tseytlin, Nucl. 
						Phys. {\bf B489}, 65 (1997),  het-th/9610172; 
  I.R. Klebanov and M. Krasnitz, Phys. Rev. {\bf D55}, R3250 (1997);
  M. Krasnitz and I. Klebanov, Phys. Rev. {\bf D56}, 2173 (1997), 
                                  hep-th/9703216; 
  M. Cvetic and F. Lasen, hep-th/9706071;
  H. W. Lee, Y. S. Myung, J. Y. Kim, Phys. Rev. {\bf D58}, 104006(1998),
        hep-th/9708099.
\bibitem{Fre98strings}
  D.Z. Freedman, Strings'98 lecture, http://www.itp.ucsb.edu/online/strings98/;
  H.Liu and A.A. Tseytlin, Phys. Rev. {\bf D59}, 086002(1999), hep-th/9807097;
  D.Z. Freedman, S.D. Mathur, A. Matusis and L. Rastelli, hep-th/9808006;
  G. Chalmers and K. Schalm, hep-th/9810051;
  H. Liu, hep-th/9811152.
\bibitem{Kog9903162}
  I.I. Kogan, hep-th/9903162.
\bibitem{Ghe9807034}
  A.M. Ghezelbash, M. Khorrami and A. Aghanohammadi, hep-th/9807034.
\bibitem{Gur93NPB535}
  V. Gurarie, Nucl. Phys. {\bf B410}, 535(1993);
  J.S. Caux, I.I. Kogan and T. Rsvelik, Nucl. Phys. {\bf B466}, 444(1996);
  M. Flohr, Int. J. Mod. Phys. {\bf A11}, 4147(1996);
    Int. J. Mod. Phys. {\bf A12}, 1943(1997);
    Nucl. Phys. {\bf B514}, 523(1998);
  I.I. Kogan and A. Lewis, Phys. Lett. {\bf B431}, 77(1998), hep-th/9802102;
  A. Lewis. Nucl. Phys. {\bf B539}, 367(1999), hep-th/9808068.
\bibitem{Vaf9804172}
  C. Vafa, hep-th/9804172.
\bibitem{Boe9806104}
  J. de Boer, hep-th/9806104.
\bibitem{Kim9812016}
  J.Y. Kim, H.W. Lee and Y.S. Myung, hep-th/9812016.
\bibitem{Kug79PTPS1}
  T. Kugo and I. Ojima, Prog. Theor. Phys. Suppl. {\bf 66}, 1(1979);
  Y.S. Myung, Y.J. Park and C. Jue, Mod. Phys. Lett. {\bf A7}, 2101(1992).
\bibitem{Kes9808037}
  E. Keski-Vakkuri, hep-th/9808037.
\bibitem{Ban93PRD1506}
  M. Banados, M. Henneaux, C. Teitelboim, and A. Zanelli, Phys. Rev. {\bf D48},
           1506(1993);
  H. W. Lee and Y. S. Myung, Phys. Rev. {\bf D58}, 104013 (1998), 
      hep-th/9804095; 
  H.W. Lee, N.J. Kim, and Y.S. Myung, hep-th/9805050.
\bibitem{Mal97PRD4975}
  J. Maldacena and A. Strominger, Phys. Rev. {\bf D56}, 4975 (1997), 
              hep-th/9702015; 
\bibitem{Gub97PRD7854}
  S.S. Gubser, Phys. Rev. {\bf D56}, 7854(1997), hep-th/9706100.
\bibitem{Bir97PLB281}
  D.B. Birmingham, I. Sachs and S. Sen, Phys. Lett. {\bf B413}, 281(1997);
  H.W. Lee, N.J. Kim, and Y.S. Myung, Phys. Rev. {\bf D58}, 084002(1998),
      hep-th/9803080.
\bibitem{Abr66}
  M. Abramowitz and I. Stegun, {\it Handbook of Mathematical
  Functions} (Academic Press, New York, 1966).
\end{references}
\end{document}